\documentclass[12pt,floatfix]{revtex4}
\usepackage{graphicx}
\usepackage{amsmath}
\usepackage{amssymb}
\usepackage[usenames,dvipsnames]{color}

\begin{document}

\title{{\it Ab initio} theory of helix$\leftrightarrow$coil phase transition}

\author{Alexander V. Yakubovich$^{\text *}$,
Ilia A. Solov'yov$^{\text *}$, Andrey V. Solov'yov\footnote{On leave
from the A.F. Ioffe Institute, St. Petersburg, Russia. E-mail:
ilia@fias.uni-frankfurt.de} and Walter Greiner}

\affiliation{Frankfurt Institute for Advanced Studies, Max von Laue
Str. 1, 60438 Frankfurt am Main, Germany}

\begin{abstract}
In this paper we suggest a theoretical method based on the
statistical mechanics for treating the
$\alpha$-helix$\leftrightarrow$random coil transition in alanine
polypeptides. We consider this process as a first-order phase
transition and develop a theory which is free of model parameters
and is based solely on fundamental physical principles. It describes
essential thermodynamical properties of the system such as heat
capacity, the phase transition temperature and others from the
analysis of the polypeptide potential energy surface calculated as a
function of two dihedral angles, responsible for the polypeptide
twisting. The suggested theory is general and with some modification
can be applied for the description of phase transitions in other
complex molecular systems (e.g. proteins, DNA, nanotubes, atomic
clusters, fullerenes).
\end{abstract}

\pacs{82.60.Fa, 87.15.He, 64.70.Nd, 64.60.-i}

\maketitle

\section{Introduction}
\label{intro}

The phase transitions in finite complex molecular systems, i.e. the
transition from a stable 3D molecular structure to a random coil
state or vice versa (also known as (un)folding process), has a long
standing history  of investigation (for review see, e.g.
\cite{Shakhnovich06,Ptizin_book,Shea01,Prabhu05}). The phase
transitions of this or similar nature occur or can be expected in
many different complex molecular systems and in nano objects, such
as polypeptides, proteins, polymers, DNA, fullerenes, nanotubes
\cite{Yakubovich06a_EPN}. They can be understood as first order
phase transitions, which are characterized by rapid growth of the
system's internal energy at a certain temperature. As a result, the
heat capacity of the system as a function of temperature acquires a
sharp maximum at the temperature of the phase transition.

In our recent paper \cite{Yakubovich06a} a novel {\it ab initio}
theoretical method for the description of phase transitions in the
mentioned molecular systems has been suggested. In particular, it
was demonstrated that in polypeptides (chains of amino acids) one
can identify specific, so-called twisting degrees of freedom
responsible for the folding dynamics of amino acid chains, i.e. for
the transition from a random coil state of the chain to its
$\alpha$-helix structure. The twisting degrees of freedom are also
sometimes referred as the torsion degrees of freedom. The essential
domain of the potential energy surface of polypeptides with respect
to these twisting degrees of freedom can be calculated and
thoroughly analyzed on the basis of {\it ab initio} methods such as
density functional theory (DFT) or Hartree-Fock method. It was shown
\cite{Yakubovich06a} that this knowledge is sufficient for the
construction of the partition function of a polypeptide chain and
thus for the development of its complete thermodynamic description,
which includes the calculation of all essential thermodynamic
variables and characteristics, e.g. free energy, heat capacity,
phase transition temperature, etc. The method has been proved to be
applicable for the description of the phase transition in
polyalanine chains of different lengths by the comparison of the
theory predictions with the results of several independent
experiments and of molecular dynamics simulations. Similar
descriptions can be developed for a large variety of complex
molecular systems.

Earlier studies of the folding process based on the statistical
mechanics principles (see
\cite{Zimm59,Gibbs59,Lifson61,Schellman58}) always contained some
empirical parameters and thus could hardly be used for {\it ab
initio} predictions of essential characteristics of the phase
transitions. Since then, the total number of papers devoted to this
problem is very large. Here we do not intend to review all of them,
but refer in this article only to those, which are related directly
to our work (for review see also
\cite{Shakhnovich06,Shea01,Prabhu05} and references therein).

The first theoretical attempt to describe the folding process of
polypeptides was done by Zimm and Bragg \cite{Zimm59}. In their work
the process of polypeptide $\alpha$-helix formation was considered
within the framework of simple two-state statistical model. This
model contains three principal parameters: (i) a constant describing
the probability of an amino acid to bond in the helix conformation
to a part of the chain being in the helical form, (ii) a special
correction factor for the initiation of helix formation (i.e. a
factor describing the probability of an amino acid to bond in the
helix conformation to an amino acid that is in the random coil
state), and (iii) the minimum number of amino acids allowed to exist
in the random coil state between two helical parts.

A different set of parameters was suggested in \cite{Gibbs59}. The
major parameters used in that paper are the energies of hydrogen
bonds in the polypeptide chain and the number of possible
conformations in the random coil state. These two parameters define
the energy and entropy differences between folded and unfolded
states of the polypeptide. In \cite{Schellman58} the factors
affecting the stability of polypeptide structures in solution were
discussed.

In \cite{Lifson61} the partition function of a polypeptide chain was
determined as a function of generalized coordinates corresponding to
the twisting degrees of freedom of the molecule's backbone. In that
paper the conditional probabilities of the occurrence of helical and
coil states of the peptide units are obtained in the form of a
$3\times3$ matrix. The eigenvalues of this matrix yield the various
molecular averages as functions of the degree of polymerization,
temperature, and molecular constants. The theoretical model
suggested in \cite{Lifson61} contained three parameters which
describe the statistical weights of three possible states of an
amino acid in a polypeptide chain: the helix state, the coil state
and the boundary state occurring at the interface between the helix
and the coil phases.

In \cite{Lifson64} another method was suggested for the derivation
of the partition function of linear-chain molecules. The partition
function was constructed on the basis of the so-called defining
sequences, being a sequence of numbers that describe the lengths of
the polypeptide parts found in different conformational states.
Therefore the defining sequence describes a certain microstate of
the system. The partition function of the system was constructed
from the partition functions of the defining sequences. To do so,
some special functions were introduced, which are called as the
sequence-generating functions.
The method suggested in \cite{Lifson64} was used in \cite{Poland66a}
for the study of helix-coil transition in polypeptides. In that
paper the conditions for the occurrence of phase transition in one
dimensional system were analyzed. In \cite{Poland66b} the kinetics
of helix-coil transition was studied within the theoretical
frameworks developed in \cite{Lifson61,Lifson64}.

In \cite{Go76,NGo69} the importance of various internal degrees of
freedom in polypeptide was discussed. The partition function of the
system was constructed within the framework of classical and quantum
mechanics.

The helix-coil transition of polypeptides was also studied in
Refs.~\cite{Ooi91,Gomez95}. In those papers general equations of
statistical physics were used to describe this transition. Those
theories contained several parameters (such as enthalpy, entropy,
free energy changes) which were fitted to represent results of
independent experimental observations.

The molecular dynamics (MD) approach, an alternative to using
statistical physics, has been widely used during the last decade for
studying structural transitions in polypeptides. Full atomistic
molecular dynamics \cite{Tobias91,Garcia02,Nymeyer03} and
Monte-Carlo based techniques \cite{Irbaeck04,Shental-Bechor05} were
used for studying alanine tripeptide \cite{Tobias91}, alanine
pentapeptide \cite{Garcia02} and alanine 21-peptide
\cite{Nymeyer03,Shental-Bechor05}. The molecular dynamics
simulations were carried out within the framework of classical
mechanics with an empirical Hamiltonian usually referred as the
forcefield. The most popular forcefields developed during recent
years are GROMOS \cite{GROMOS}, AMBER \cite{AMBER} and CHARMM
\cite{CHARMM}.

During the last years molecular dynamics was also widely applied for
studying the folding process of small proteins
\cite{Chen05,Duan98,Liwo05,Ding02,Pande02,Irbaeck03}. Such
simulations became possible relatively recently due to modern
computer powers. However, it is still not feasible to perform
molecular dynamics simulations of the folding process of large
proteins \cite{Shakhnovich06} because the characteristic timescale
of this process varies from micro seconds to minutes
\cite{Kubelka04,Lipman03}, being several orders of magnitude larger
than the time of possible molecular dynamics simulations.

Another molecular dynamics approach for studying the protein folding
problem was suggested in \cite{He98a,He98b}. In these papers the
dynamics of the macromolecule was considered in the phase space of
torsional degrees of freedom.

Stochastic treatment of helix-coil transition in polypeptides was
performed in \cite{Fujita81,Cardenas03}. In \cite{Fujita81} the
application of correlated random walk theory for polypeptides was
analyzed. In \cite{Cardenas03} an atomistic simulation of helix
formation with the stochastic difference equation was performed.

The helix-coil transition of polypeptides has also been extensively
studied experimentally
\cite{Scholtz91,Lednev01,Thompson97,Williams96}. In \cite{Scholtz91}
the enthalpy change accompanying the $\alpha$-helix to coil
transition has been determined calorimetrically for a 50-residue
Ac-Y(AEAAKA)$_8$F-NH$_2$ peptide that contains primarily alanine.
The dependence of the heat capacity of the polypeptide on
temperature was measured with the use of differential scanning
calorimetry method. In \cite{Lednev01,Thompson97} the experiments
were performed for A$_5$(A$_3$RA)$_3$A and
MABA-A$_5$-(AAARA)$_3$-A-NH$_2$ alanine-rich peptides consisting of
$21$ amino acids by means of UV resonance Raman spectroscopy and by
circular dichroism, respectively. The dependence of helicity on
temperature was recorded. Kinetics of the helix-coil transition of
21 residue Suc-AAAAA-(AAARA)$_3$A-NH$_2$ alanine based polypeptide
was studied in \cite{Williams96} by means of infrared spectroscopy.

Previous attempts to describe the helix-coil transition in
polypeptide chains within the framework of statistical physics were
based on the models suggested in the sixties
\cite{Zimm59,Gibbs59,Lifson61,Schellman58}, where the general
formalism for the construction of the partition function of
polypeptides was suggested.
Earlier theories always included several parameters in the partition
function making it parameter dependent. The methods suggested in
\cite{Zimm59,Gibbs59,Lifson61,Schellman58} were widely used for the
description of the helix-coil transition in polypeptide chains (see
Refs.~\cite{Kromhout01,Chakrabartty94,Shakhnovich06,Ptizin_book,Shea01,Scheraga70,Scheraga02,Shental-Bechor05}).
The dependance of the thermodynamic characteristics of the
$\alpha$-helix$\leftrightarrow$random coil phase transition in
polypeptides on model parameters, used for the partition function
construction, was thoroughly analysed (see papers cited above). Some
attempts were made to obtain these parameters from experimental
observations and from the theoretical calculations. In
\cite{Wojcik90} the parameters of the Zimm and Bragg theory
\cite{Zimm59} were deduced from the optical rotatory dispersion and
circular dichroism measurements on poly(L-cystine) in water at
neutral pH.

The first attempts to evaluate the parameters of the Zimm-Bragg
theory theoretically were performed in \cite{Scheraga70}. In that
paper a semi-empirical potential \cite{Scott66,Ooi67} was used to
describe the conformational dynamics of the polypeptide. The
potential suggested in these papers is similar to the modern
forcefields \cite{GROMOS,AMBER,CHARMM}, but treats the structure of
a polypeptide in a simplified way by neglecting some of the hydrogen
atoms in the polypeptide and making minimal assumptions about the
hybridization of atoms. The potential used in \cite{Scott66,Ooi67}
can be considered as one of the first (if not the first) forcefields
suggested. With its use in \cite{Scheraga70} the parameters of the
Zimm-Bragg theory were calculated and the temperature of the
helix-coil transition in polypeptide chain was established. In that
paper the partition function was constructed and evaluated within a
matrix approach developed in \cite{Lifson61}

The parameters of the Zimm-Bragg theory were also calculated by
means of molecular dynamics simulation \cite{Wang95}. A peptide
growth simulation method was introduced, which allowed the
generation of dynamic models of polypeptide chains in $\alpha$-
helix or random coil conformations. With this method the Zimm-Bragg
parameters for helix initiation and helix growth have been
calculated.

In the present paper we describe an alternative theoretical approach
based on the statistical mechanics for treating the
$\alpha$-helix$\leftrightarrow$random coil phase transition in
alanine polypeptides. The suggested method is a further development
of the method suggested in \cite{Yakubovich06a,Yakubovich06a_EPN},
which is based on the construction of a parameter-free partition
function for a system experiencing a phase transition. All the
necessary information for the construction of such a partition
function can be calculated on the basis of {\it ab initio} DFT,
combined with molecular mechanics theories. Comparison of the
results of this method with the results of molecular dynamics
simulations (see following paper \cite{Yakubovich07_following})
allows one to establish the accuracy of the new approach for
sufficiently large molecular systems and then to extend the
description to the larger molecular objects, which is especially
essential in those cases when molecular dynamics simulations are
hardly possible because of computer power limitations.

We note that the suggested method is considered as an efficient
novel alternative to the existing theoretical approaches for the
study of helix-coil transitions in polypeptides since it does not
contain any model parameters and gives a universal recipe for the
construction of the partition function in complex molecular systems.
The partition function of the polypeptide is constructed based on a
minimal number of assumptions about the system which are different
from those used in earlier theories. It includes all essential
physical contributions needed for the description of the helix-coil
transition in polypeptides. Therefore the final expression for the
partition function obtained within the framework of our theory is
different from the ones suggested earlier.

In this paper we present in detail the theoretical method for the
study of $\alpha$-helix$\leftrightarrow$random coil phase
transitions in polypeptides, while in the following paper
\cite{Yakubovich07_following} we report the results of numerical
simulations of this process.

\section{Statistical model for the $\alpha$-helix$\leftrightarrow$random coil phase transition}
\label{theory}

Let us consider a polypeptide, consisting of $n$ amino acids. The
polypeptide can be found in one of its numerous isomeric states that
have different energies. A group of isomeric states with similar
characteristic physical properties is called a {\it phase state} of
the polypeptide. Thus, a regular bounded $\alpha$-helix state
corresponds to one phase state of the polypeptide, while all
possible unbounded random conformations can be denoted as the random
coil phase state.

\begin{figure}[h]
\includegraphics[scale=0.82,clip]{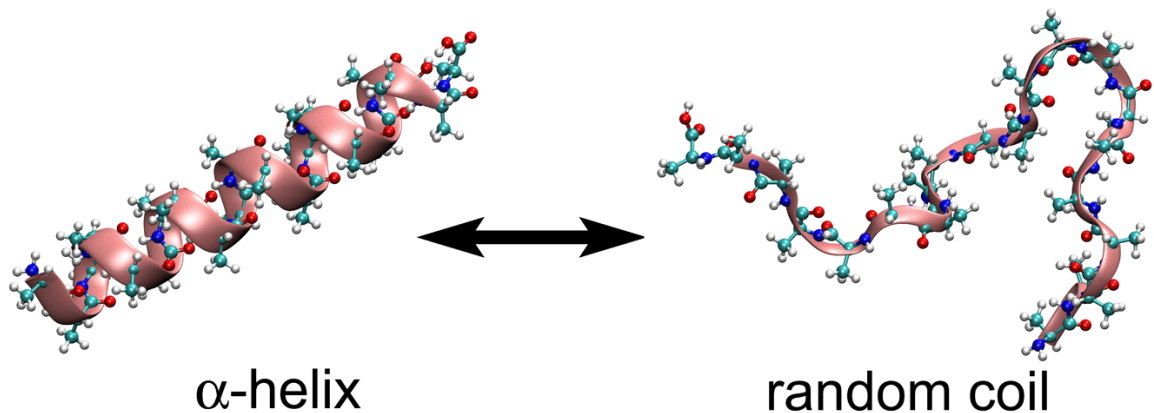}
\caption{The characteristic structural change of alanine polypeptide
experiencing an $\alpha$-helix$\leftrightarrow$random coil phase
transition.} \label{fg:helix-coil}
\end{figure}

The {\it phase transition} is the transformation of the polypeptide
from one phase state to another, i.e. the transition from a regular
$\alpha$-helix conformation to a group of unbounded random
conformations. The characteristic structural change of alanine
polypeptide experiencing an $\alpha$-helix$\leftrightarrow$random
coil phase transition is shown in Fig. \ref{fg:helix-coil}. In this
figure we show only one characteristic conformation of the
polypeptide in the random coil state, while there exist about
$10^{30}$ different conformations of 21 alanine polypeptide (see
\cite{Yakubovich06a} for more details).

The phase transition can either be of the first or of the second
order. The first order phase transition is characterized by an
abrupt change of the internal energy of the system with respect to
its temperature. In the first order phase transition the system
either absorbs or releases a fixed amount of energy while the heat
capacity as a function of temperature has a pronounced peak
\cite{Ptizin_book}. We study the manifestation of these features for
alanine polypeptide chains of different lengths.

\subsection{Hamiltonian of a polypeptide chain}

To study thermodynamic properties of the system one needs to
investigate its potential energy surface with respect to all degrees
of freedom. There are a number of different methods for calculating
the energy of many-body systems. The most accurate approaches are
based on solving the Schr{\"o}dinger equation. These approaches are
usually referred as {\it ab initio} methods since they involve a
minimum number of assumptions about the system.


For complex molecular systems {\it ab initio} calculations require
significant computer power. Depending on the method, the
computational cost of such calculations grows as $N^2$ or even $N^8$
\cite{Gaussian98_man}, where $N$ is the number of particles in the
system. The size of molecular system which can be described using
{\it ab initio} methods is therefore limited, and such methods can
hardly be used for the description of large biological molecules or
systems.

For the description of macromolecular systems, such as polypeptides
and proteins, efficient model approaches are necessary. One of the
most common tools for the description of macromolecules is based on
the so-called molecular mechanics potential, which reads as

\begin{eqnarray}
\nonumber U= \sum_{i=1}^{N_b} k_i^{b}(r_i-r_i^0)^2+\sum_{i=1}^{N_a}
k_i^{a}(\theta_i-\theta_i^0)^2+
\sum_{i=1}^{N_d} k_i^{d}\left[1+\cos(n_i\phi_i+\delta_i)\right]+\\
\sum_{i=1}^{N_{id}}k_i^{id}(S_i-S_i^0)^2+
\sum_{{i,j=1}\atop{i<j}}^{N}4\epsilon_{ij}\left[\left(\frac{\sigma_{ij}}{r_{ij}}\right)^{12}-\left(\frac{\sigma_{ij}}{r_{ij}}\right)^6\right]+
\sum_{{i,j=1}\atop{i<j}}^{N}\frac{q_iq_j}{r_{ij}}. \label{mol_mech}
\end{eqnarray}

\noindent Here the first four terms describe the potential energy
with respect to variation of distances, angles, dihedral angles and
improper dihedral angles between two, three and four neighboring
atoms respectively. The last two terms describe the van der Waals
and Coulomb interaction respectively. The summation in the first
term goes over all topologically defined bonds in the system, in the
second over all topologically defined angles, and in the third over
all topologically defined dihedral angles and in the fourth over all
topologically defined improper dihedral angles. The total number of
bonds, angles, dihedral angles and improper dihedral angles are
$N_b$, $N_a$, $N_d$ and $N_{id}$ respectively. $N$ is the total
number of atoms in the system. $k_i^{b}$, $k_i^{a}$, $k_i^{d}$ and
$k_i^{id}$ in (\ref{mol_mech}) are the stiffness parameters of the
corresponding energy terms. $r_i^0$, $\theta_i^0$ and $S_i^0$ are
the equilibrium values of bonds, angles and improper dihedral
angles. $n_i$ and $\delta_i$ are the number of possible stable
torsion conformations and the initial torsion phase.
$\epsilon_{ij}$, $\sigma_{ij}$ and $q_i$ are the van der Waals
parameters and the charges of atoms in the system.

Parameters $k_i^{b}$, $k_i^{a}$, $k_i^{d}$, $k_i^{id}$, $r_i^0$,
$\theta_i^0$, $S_i^0$, $n_i$, $\delta_i$, $\epsilon_{ij}$,
$\sigma_{ij}$ and $q_i$ are derived from experimental measurements
of crystallographic structures, infrared spectra or on the basis of
quantum mechanical calculations for small systems (see
\cite{GROMOS,AMBER,CHARMM} and references therein). The independent
variables in (\ref{mol_mech}) are $r_i$, $\theta_i$, $\phi_i$ and
$S_i$.

\begin{figure}[h]
\includegraphics[scale=0.8,clip]{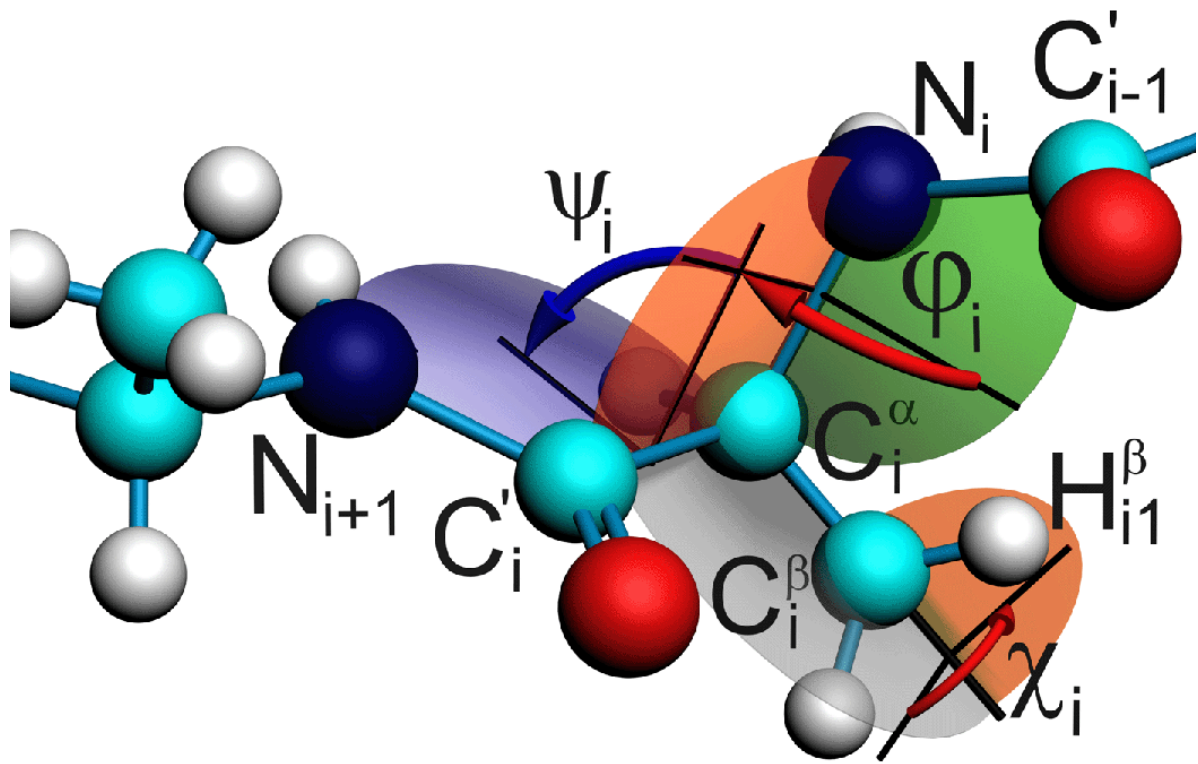}
\caption{Dihedral angles $\varphi$ and $\psi$ used for
characterization of the secondary structure of a polypeptide chain.
The dihedral angle $\chi_i$ characterizes the rotation of the side
radical along the $C_{i}^{\alpha}-C_i^{\beta}$ bond.}
\label{fg:angle_def}
\end{figure}

Note, that the terms corresponding to the variations of distances,
angles and improper dihedral angles in (\ref{mol_mech}) describe the
motion of the molecule within the harmonic approximation which is
reasonable only at low temperatures. The potential energy
corresponding to torsion degrees of freedom is usually assumed to be
periodic (see equation (\ref{mol_mech})) because several stable
conformations of the molecule with respect to these degrees of
freedom are possible
\cite{GROMOS,AMBER,CHARMM,Yakubovich06b,Yakubovich06c,ISolovyov06b,ISolovyov06c}.
The torsion degrees of freedom are also referred as the twisting
degrees of freedom
\cite{Yakubovich06b,Yakubovich06c,ISolovyov06b,ISolovyov06c}. The
most important twisting degrees of freedom for the description of a
helix-coil transition in polypeptides are the twisting degrees of
freedom along the backbone of the polypeptide
\cite{Yakubovich06a,Yakubovich06a_EPN,He98a,He98b}. These degrees of
freedom are defined for each amino acid of the polypeptide except
for the boundary ones and are described by two dihedral angles
$\varphi_i$ and $\psi_i$ (see Fig. \ref{fg:angle_def})

Both angles are defined by four neighboring atoms in the polypeptide
chain. The angle $\varphi_i$ is defined as the dihedral angle
between the planes formed by the atoms
($C_{i-1}^{'}-N_{i}-C_i^{\alpha}$) and
($N_{i}-C_i^{\alpha}-C_i^{'}$). The angle $\psi_i$ is defined as the
dihedral angle between the ($N_{i}-C_{i}^{\alpha}-C_i^{'}$) and
($C_{i}^{\alpha}-C_i^{'}-N_{i+1}$) planes. The atoms are numbered
from the NH$_2$- terminal of the polypeptide. The angles $\varphi_i$
and $\psi_i$ take all possible values within the interval
[$-180^{\circ}$;$180^{\circ}$]. For the unambiguous definition the
angles $\varphi_i$ and $\psi_i$ are counted clockwise, if one looks
on the molecule from its NH$_2$- terminal (see Fig.
\ref{fg:angle_def}). This way of angle counting is the most commonly
used
\cite{Rubin04,Yakubovich06b,Yakubovich06c,ISolovyov06b,ISolovyov06c}.


A Hamiltonian function of a polypeptide chain is constructed as a
sum of the potential, kinetic and vibrational energy terms. For a
polypeptide chain in a particular conformational state $j$
consisting of $n$ amino acids and $N$ atoms we obtain:

\begin{equation}
H_{j}=\frac{{\bf P^{2}}}{2 M}+\frac{1}{2}\left(I_{1}
^{(j)}\Omega_{1}^2+I_{2}^{(j)}\Omega_2^2+I_{3}^{(j)}\Omega_{3
}^2\right)+\sum_{i=1}^{3N-6}\frac{p_i^2}{2 m_i}+U(\{x\}),
 \label{ham_func_1}
\end{equation}

\noindent where ${\bf P}$, $M$, $I_{1,2,3}^{(j)}$, $\Omega_{1,2,3}$,
are the momentum of the whole polypeptide, its mass, its three main
momenta of inertia,  and its rotational frequencies. $p_i$, $x_i$
and $m_i$ are the momentum, the coordinate and the generalized mass
describing the motion of the system along the $i$-th degree of
freedom. $U(\{x\})$ is the potential energy of the system, being the
function of all atomic coordinates in the system.

One can group all degrees of freedom in a polypeptide in the two
classes: "stiff" and "soft" degrees of freedom. We call the degrees
of freedom corresponding to the variation of bond lengths, angles
and improper dihedral angles (see Fig. \ref{fg:angle_def}) as
"stiff", while degrees of freedom corresponding to the angles
$\varphi_i$ and $\psi_i$ are classified as "soft" degrees of
freedom. The "stiff" degrees of freedom can be treated within the
harmonic approximation because the energies needed for a noticeable
change of the system structure with respect to these degrees of
freedom are about several eV which is significantly larger than the
characteristic thermal energy of the system at room temperature
being on the order of $0.026$ eV
\cite{ISolovyov06a,ISolovyov06b,ISolovyov06c,GROMOS,AMBER,CHARMM}.

The Hamiltonian of the polypeptide can be rewritten in terms of the
"soft" and "stiff" degrees of freedom. Transforming the set of
cartesian coordinates $\{x\}$ to a set of generalized coordinates
$\{q\}$, corresponding to the "soft" and "stiff" degrees of freedom
one obtains:

\begin{eqnarray}
\nonumber H_{j}=\frac{{\bf P^{2}}}{2 M}+\frac{1}{2}\left(I_{1}
^{(j)}\Omega_{1}^2+I_{2}^{(j)}\Omega_2^2+I_{3}^{(j)}\Omega_{3
}^2\right)
+\sum_{i=1}^{l_s}\sum_{j=l}^{l_s}g_{ij}\frac{p_i^s p_j^s}{2}\\
+\sum_{i=1}^{l_s}\sum_{j=l_s+1}^{l_h}g_{ij}p_i^s p_j^h
+\sum_{i=l_s+1}^{l_h}\sum_{j=l_s+1}^{l_h}g_{ij}\frac{p_i^h p_j^h}{2}
+U(\{q^s\},\{q^h\}), \label{eq:ham_func_gen_coor}
\end{eqnarray}

\noindent where $q^s$ and $q^h$ are the generalized coordinates
corresponding to the "soft" and "stiff" degrees of freedom, and
$p^s$ and $p^h$ are the corresponding generalized momenta. $l_s$ and
$l_h$ is the number of the "soft" and "stiff" degrees of freedom in
the system, satisfying the relation $3N-6=l_s+l_h$.
$U(\{q^s\},\{q^h\})$ in Eq.~(\ref{eq:ham_func_gen_coor}) is the
potential energy of the system as a function of the "soft" and
"stiff" degrees of freedom. $1/g_{ij}$ has a meaning of the
generalized mass, while $g_{ij}$ is defined as follows:

\begin{eqnarray}
g_{ij}=\sum_{\lambda=1}^{3N-6}\frac{1}{m_\lambda}\frac{\partial
q_i}{\partial x_{\lambda}}\frac{\partial q_j}{\partial x_{\lambda}}
\label{g_ij}.
\end{eqnarray}

\noindent Here $x_\lambda$ and $m_\lambda$ are the generalized
coordinate in the cartesian space and the generalized mass of the
system, corresponding to the degree of freedom with index $\lambda$.
$q_i$ and $q_j$ denote the "soft" or the "stiff" generalized
coordinate in the transformed space.

The motion of the system with respect to its "soft" and "hard"
degrees of freedom occurs on the different time scales as was
discussed in \cite{NGo69}. The typical oscillation frequency
corresponding to the "soft" degrees of freedom is on the order of
100 cm$^{-1}$, while for the "stiff" degrees of freedom it is more
than 1000 cm$^{-1}$ \cite{NGo69}. Thus the motion of the system with
respect to the "soft" degrees of freedom is uncoupled from the
motion of the system with respect to the "stiff" degrees of freedom.
Therefore the fifth term in Eq.~(\ref{eq:ham_func_gen_coor}), which
describes the kinetic energy of the "stiff" motions in the
polypeptide can be diagonalized. The corresponding set of
coordinates $\{\tilde{q}^s\}$ describes the normal vibration modes
in the "stiff" subsystem:

\begin{eqnarray}
\nonumber H_{j}=\frac{{\bf P^{2}}}{2 M}+\frac{1}{2}\left(I_{1}
^{(j)}\Omega_{1}^2+I_{2}^{(j)}\Omega_2^2+I_{3}^{(j)}\Omega_{3
}^2\right)+\sum_{i=1}^{l_h}\left(\frac{\left(\tilde{p}^h_i\right)^2}{2\mu_i^h}+\frac{\mu_i^h\omega_i^2\left(\tilde{q}^h_i\right)^2}{2}\right)\\
+\sum_{i=1}^{l_s}\sum_{j=1}^{l_s}g_{ij}\frac{p_i^s p_j^s}{2}
+U(\{\chi\})+U(\{\varphi,\psi\}). \label{ham_func_2}
\end{eqnarray}

\noindent Here $\omega_i$ and $\mu_i^h$ are the frequency of the
$i$-th "stiff" normal vibrational mode and the corresponding
generalized mass. Note, that the fourth term in
Eq.~(\ref{eq:ham_func_gen_coor}) vanishes if the "soft" and the
"stiff" degrees of freedom are uncoupled. The last two terms in
Eq.~(\ref{ham_func_2}) describe the potential energy of the system
in respect to the "soft" degrees of freedom. For every amino acid
there are at least two "soft" degrees of freedom, corresponding to
the angles $\varphi_i$ and $\psi_i$ (see Fig. \ref{fg:angle_def}).
Some additional "soft" degrees of freedom involve the rotation of
the side radicals in amino acids. A typical example is the angle
$\chi_i$, which describes the twisting of the side chain radical
along the $C^{\alpha}_i-C^{\beta}_i$ bond (see
Fig.~\ref{fg:angle_def}). The angle $\chi_i$ is defined as the
dihedral angle between the planes formed by the atoms
($C_{i}^{'}-C_i^{\alpha}-C_i^{\beta}$) and by the bonds
$C_i^{\alpha}-C_i^{\beta}$ and $C_i^{\beta}-H_{i1}^{\beta}$. Note,
that the notations $\chi$, $\varphi$ and $\psi$ are used for the
simplicity and for the further explanation of our theory. The set of
these dihedral angles builds up the set of "soft" degrees of freedom
of the polypeptide: $\{q^s\}\equiv\{\chi,\varphi,\psi\}$.

Note that generalized masses $1/g_{ij}$ depend on the choice of the
generalized coordinates in the system. However this dependence can
be neglected if the system is considered in the vicinity of its
equilibrium state. In this case the motion of the polypeptide with
respect to the "soft" degrees of freedom can be considered as the
motion of the system of coupled nonlinear oscillators. In the
vicinity of the system's equilibrium state the generalized mass can
be written as:

\begin{equation}
\frac{1}{g_{ij}}=\frac{1}{g_{ij}\left(\{q^s_{i_0}\}\right)}+\sum_{k=1}^{l_s}\left.\frac{\partial\left(1/g_{ij}\right)}{\partial
q^s_k}\right|_{q_k^s=q_{k_0}^s}\left(q_k^s-q_{k_0}^s\right),
\label{eq:gen_mass}
\end{equation}

\noindent where $q_{k_0}^s$ denotes the value of the $k$-th "soft"
degree of freedom at the equilibrium position. The second term in
Eq.~(\ref{eq:gen_mass}) describes the dependence of the generalized
mass on coordinates and can be neglected if the system is in the
vicinity of its equilibrium. All the information about the
nonlinearity of the oscillations is contained in the potential
energy functions $U(\{\chi\})$ and $U(\{\varphi,\psi\})$ in
Eq.~(\ref{ham_func_2}).

The validity of the coordinate-independent mass approximation was
also discussed in Ref.~\cite{NGo69}. In the present paper we do not
account for the coordinate dependence of the generalized masses,
$g_{ij}$, and leave this question open for further investigation.

\subsection{Partition function}
\label{sec:part_func}

The partition function of the polypeptide is constructed within the
framework of classical mechanics. We consider the classical
partition function because in our following paper
\cite{Yakubovich07_following} we have treated the polypeptide
classically. However the presented formalism can be easily
generalized for the quantum mechanical description of the system.

All thermodynamic properties of a system are determined by its
partition function, which can be expressed via the system's
Hamiltonian in the following form \cite{Landau05}:

\begin{equation}
{\mathbb Z}=\int\exp\left(-\frac{H}{kT}\right)d\Gamma,
 \label{par_gen_1}
\end{equation}

\noindent where $H$ is the Hamiltonian of the system, $k$ and $T$
are the Boltzmann constant and the temperature respectively and
$d\Gamma$ is an element of the phase space. Substituting
(\ref{ham_func_2}) into (\ref{par_gen_1}) one obtains an expression
for the partition function of a polypeptide in a particular
conformational state $j$. Thus, the partition function of the system
can be factored as follows:

\begin{equation}
{\mathbb Z}=\frac{1}{\left(2\pi\hbar\right)^{3N}}Z_1\cdot Z_2\cdot
Z_3\cdot Z_4\cdot Z_5, \label{eq:Znone}
\end{equation}

\noindent where

\begin{eqnarray}
\nonumber Z_1&=&\int\exp\left(\frac{1}{kT}\left[-\frac{{\bf
P^{2}}}{2 M}-\left(\frac{{\cal M}_1^2}{2I_{1} ^{(j)}}+\frac{{\cal
M}_2^2}{2I_{2} ^{(j)}}+\frac{{\cal M}_3^2}{2I_{3} ^{(j)}}\right)\right]\right){\rm d^3}P\cdot{\rm d^3}Q\cdot {\rm d^3}{\cal M}\cdot{\rm d^3}\Phi=\\
&&=64\pi^5V_{j}M^{3/2}\sqrt{I_{j}^{(1)}I_{j}^{(2)}I_{j}^{(3)}}(kT)^3\label{eq:term1}
\end{eqnarray}

\begin{eqnarray}
 \label{eq:term2}
Z_2&=&\int\exp\left(-\frac{1}{kT}\sum_{i=1}^{l_h}\left(\frac{\left(\tilde{p}^h_i\right)^2}{2\mu_i^h}+\frac{\mu_i^h\omega_i^2\left(\tilde{q}^h_i\right)^2}{2}\right)\right){\rm
d^{l_h}}\tilde{p}^h \cdot{\rm d^{l_h}}\tilde{q}^h=\frac{\left(2\pi k
T\right)^{l_h}}{\prod_{i=1}^{l_h}\omega_i},
\end{eqnarray}

\begin{eqnarray}
 \label{eq:term3}
Z_3&=&\int\exp\left(-\frac{1}{kT}\sum_{i=1}^{l_s}\frac{\left(\tilde{p}^s_i\right)^2}{2\mu_i^s}\right){\rm
d^{l_s}}\tilde{p}^s =\sqrt{\left(2\pi k
T\right)}^{l_s}\prod_{i=1}^{l_s}\sqrt{\mu_i^s},
\end{eqnarray}

\begin{eqnarray}
 \label{eq:term4}
Z_4&=&\int\exp\left(-\frac{U(\{\tilde{\chi}\})}{kT}\right){\rm
d^{l_\chi}}\tilde{\chi}^s,
\end{eqnarray}

\begin{eqnarray}
 \label{eq:term5}
Z_5&=&\frac{1}{\left(2\pi\hbar\right)^{(l_\varphi+l_\psi)/2}}\int\exp\left(-\frac{U(\{\tilde{\varphi},\tilde{\psi}\})}{kT}\right){\rm
d^{l_\varphi}}\tilde{\varphi}^s \cdot{\rm d^{l_\psi}}\tilde{\psi}^s.
\end{eqnarray}

\noindent $Z_1$, Eq.~(\ref{eq:term1}), describes the contribution to
the partition function originating from the motion of the
polypeptide as a rigid body. Here $V_{j}$ is the specific volume of
the polypeptide in conformational state $j$ and ${\cal M}$ is the
angular momenta of the polypeptide. $Z_2$, Eq.~(\ref{eq:term2}),
accounts for the "stiff" degrees of freedom in the polypeptide.
$Z_3$, Eq.~(\ref{eq:term3}), describes the contribution of the
kinetic energy of the "soft" degrees of freedom to the partition
function. $Z_4$, Eq.~(\ref{eq:term4}), and $Z_5$,
Eq.~(\ref{eq:term5}), describe the contribution of the potential
energy of the "soft" degrees of freedom to the partition function.
Integrating over the phase space in
Eqs.~(\ref{eq:term1})-(\ref{eq:term5}) is performed over generalized
coordinates and momentum space.

For the derivation of Eqs.~(\ref{eq:term3})-(\ref{eq:term5}) we have
diagonalized the quadratic form of the generalized momenta
corresponding to the "soft" degrees of freedom in
Eq.~(\ref{ham_func_2}) and made a transformation
$q_i^s\rightarrow{\tilde{q}}_i^s$,
$p_i^s\rightarrow{\tilde{p}}_i^s$. In Eq.~(\ref{eq:term3}),
$\mu_i^s$ is the generalized mass of the $i$-th "soft" normal
vibration mode, being related to $g_{ij}$ in Eq.~(\ref{g_ij}).
 ${\tilde{\chi}}$,${\tilde{\varphi}}$ and ${\tilde{\psi}}$ in
Eqs.~(\ref{eq:term4})-(\ref{eq:term5}) denote the "soft" twisting
degrees of freedom, which have been transformed accordingly. Note
that $\tilde{q}_i^s$ and $\tilde{p}_i^s$ are canonical conjugated
coordinates. $l_\chi$, $l_\varphi$ and $l_\psi$ in
Eqs.~(\ref{eq:term4})-(\ref{eq:term5}) is the number of the $\chi$,
$\varphi$ and $\psi$ degrees of freedom in the system. Note, that
$l_s=l_\chi+l_\varphi+l_\psi$.

Integrals in Eqs.~(\ref{eq:term1})-(\ref{eq:term3}) can be evaluated
analytically, while for the integration over the angles $\chi$,
$\varphi$ and $\psi$ in Eqs.~(\ref{eq:term4})-(\ref{eq:term5}) the
knowledge of the exact potential energy surface of the polypeptide
is necessary. However the potential energy of the polypeptide
corresponding to the twisting degrees of freedom $\chi$ does not
depend on the conformation of the polypeptide in case of neutral
non-polar radicals in simple amino acids (i.e. alanine, glycine)
\cite{NGo69}. Thus, the twisting degrees of freedom corresponding to
the variations of angles $\chi$ have a minor influence on the
$\alpha$-helix$\leftrightarrow$random coil phase transition. The
potential energy of the polypeptide in respect to these degrees of
freedom is well described by the following function, as follows from
the molecular mechanics potential Eq.~(\ref{mol_mech}):

\begin{equation}
U(\chi_i)=k_{\chi_i}\left[1+\cos\left(3\chi_i\right)\right],
\label{eq:chi}
\end{equation}

\noindent where $k_{\chi_i}$ is the stiffness parameter of the
potential. Since $k_{\chi_i}=k_{\chi}$, substituting
Eq.~(\ref{eq:chi}) into Eq.~(\ref{eq:term4}) and integrating over
$2\pi$ one obtains:

\begin{eqnarray}
 \label{eq:term4mod}
Z_4&=&\left[2\pi\exp\left(-\frac{k_{\chi}}{kT}\right){\rm
I_0}\left(\frac{k_{\chi}}{kT}\right)\right]^{l_\chi}=(2\pi)^{l_\chi}
B(kT),
\end{eqnarray}

\noindent where ${\rm I_0}(x)$ is the the modified Bessel function
of the first kind, and
$B(kT)=\left[\exp\left(-\frac{k_{\chi}}{kT}\right){\rm
I_0}\left(\frac{k_{\chi}}{kT}\right)\right]^{l_\chi}$.

Substituting $Z_{1}$-$Z_{5}$ into Eq.~(\ref{eq:Znone}) one obtains
the expression for the partition function of a polypeptide in a
particular conformational state $j$:

\begin{eqnarray}
\nonumber Z_{j}&=&\left[ \frac{V_{j}\cdot
M^{3/2}\cdot\sqrt{I_{j}^{(1)}I_{j}^{(2)}I_{j}^{(3)}}
\prod_{i=1}^{l_s}\sqrt{\mu_i^s}}{(2\pi)^{\frac{l_s}{2}-l_\chi}\pi
    \hbar^{3N}\prod_{i=1}^{l_h}\omega_i}\right]B(kT)\cdot(kT)^{3N-3-\frac{l_s}{2}}\\
    \nonumber
    &&\int_{-\pi}^{\pi }\ldots \int_{-\pi}^{\pi}
     e^{-\frac{U(\{\varphi,\psi\})}{kT}}{\rm d}\varphi_1\ldots {\rm d}\varphi_n\ {\rm d}\psi_1\ldots {\rm d}\psi_n=\\
&=&A_{j}\cdot B(kT)\cdot (kT)^{3N-3-\frac{l_s}{2}}\int_{-\pi}^{\pi
}\ldots \int_{-\pi}^{ \pi}
     e^{-\frac{U(\{\varphi,\psi\})}{kT}}{\rm d}\varphi_1\ldots {\rm d}\varphi_n\ {\rm d}\psi_1\ldots {\rm d}\psi_n, \label{part_func_1}
\end{eqnarray}

\noindent $A_{j}$ denotes the factor in the square brackets. Note,
that generalized masses $\mu_i^h$ are reduced during the integration
and do not enter into the expression of the partition function.

Since a polypeptide exist in different conformational states, one
needs to sum over the contributions of all possible conformations
$Z_j$ in order to calculate the complete partition function of the
polypeptide. For an ensemble of ${\cal N}$ noninteracting
polypeptides the partition function reads as

\begin{eqnarray}
\nonumber  {\mathbb Z}=\left(\sum_{j=1}^{\xi}Z_j\right)^{\cal
N}&=&\left(B(kT)\cdot(kT)^{3N-3-\frac{l_s}{2}}\sum_{j=1}^{\xi}A_j\right.\\
&&\left.\cdot\int_{-\pi}^{\pi }\ldots \int_{-\pi}^{ \pi}
e^{-\frac{U(\{\varphi,\psi\})}{kT}}{\rm d}\varphi_1\ldots {\rm
d}\varphi_n\ {\rm d}\psi_1\ldots {\rm d}\psi_n\right)^{\cal N},
\label{Z_step1}
\end{eqnarray}

\noindent where $Z_j$ is defined in (\ref{part_func_1}) and $\xi$ is
the total number of possible conformations in a polypeptide.
Equation (\ref{Z_step1}) has been derived with a minimum number of
assumptions about the system. It is general, however, its use for a
particular molecular systems is not so straightforward. Expression
(\ref{Z_step1}) can be further simplified, if one makes additional
assumptions about the structure of the system.

For the sake of simplicity, we write further equations for only one
polypeptide instead of ${\cal N}$. Generalization for the case of
${\cal N}$ statistically independent polypeptides can always be done
according to (\ref{Z_step1}).

One can expect that the factors $A_j$ in (\ref{Z_step1}) depend on
the chosen conformation of the polypeptide. However, due to the fact
that the values of specific volumes, momenta of inertia and
frequencies of normal vibration modes of the polypeptide in
different conformations are expected to be close
\cite{Krimm80,Yakubovich06a}, the values of $A_j$ in all these
conformations can be considered as equal, at least in the zero order
approximation. Thus $A_j\equiv A$.

The amino acids can be treated as statistically independent in any
conformation of the polypeptide. This fact is not obvious and it was
not systematically investigated so far. The statistical independence
of small neutral non-polar amino acids (alanine, glycine, etc) in a
polypeptide was studied in \cite{Rubin04} with the use of
time-correlation functions between different amino acids. In our
following paper \cite{Yakubovich07_following}, we address this
question for alanine polypeptides and determine the degree to which
amino acids in the polypeptide can be treated as statistically
independent.

With the assumptions made, the partition function of polypeptide
reduces to:

\begin{equation}
{\mathbb Z}=A\cdot
B(kT)\cdot(kT)^{3N-3-\frac{l_s}{2}}\sum_{j=1}^{\xi}\prod_{i=1}^{n}\int_{-\pi}^{\pi
}\int_{-\pi}^{ \pi}
\exp\left({-\frac{\epsilon_i^{(j)}(\varphi,\psi)}{kT}}\right){\rm
d}\varphi {\rm d}\psi, \label{Z_step2}
\end{equation}

\noindent where $\epsilon_i^{(j)}(\varphi,\psi)$ is the potential
energy of $i$-th amino acid in the polypeptide, being in one of its
$\xi$ conformations denoted with $j$. The potential energy of the
amino acid is calculated as a function of its twisting degrees of
freedom $\varphi$ and $\psi$.

In equation (\ref{Z_step2}) the partition function is summed over
all conformations of the polypeptide. However, in the case of the
$\alpha$-helix to random coil transition of the polypeptide, the
summation over the polypeptide conformations has to be performed
only over the conformations involved in the transition.

Note that Eq.~(\ref{Z_step2}) is rather general and can be used for
the description of the folding process in proteins. Indeed, the
partition function in Eq.~(\ref{Z_step2}) is determined by the
potential energy surfaces of amino acid in the native state of a
protein and in the random coil conformation. The potential energy
surfaces can be calculated on the basis of {\it ab initio} DFT,
combined with molecular mechanics theories as demonstrated in
\cite{Yakubovich06a,Yakubovich06a_EPN} and in the following paper
\cite{Yakubovich07_following}. For a protein, which has 20 different
amino acids it is necessary to calculate at least 40 different
potential energy surfaces, while for the study of folding of
polypeptide consisting of the identical amino acids a single
potential energy surface describes the transition.

Further simplifications of the partition function (\ref{Z_step2})
for polypeptide consisting of the identical amino acids can be
achieved if one assumes that each amino acid in the polypeptide can
occupy two states only, below referred as the {\it bounded} and {\it
unbounded states}. The amino acid is considered to be in the bounded
state when it forms one hydrogen bond with the neighboring amino
acids. In the unbounded state amino acids do not have hydrogen
bonds. When the $\alpha$-helix is formed, all amino acids are in the
bounded state, while in the case of random coil all amino acids
occupy the unbounded states.

All possible conformations of the polypeptide experiencing in the
course of the $\alpha$-helix$\leftrightarrow$random coil phase
transition can be divided in three different groups:

\begin{itemize}

\item[I.] completely folded state of the polypeptide ($\alpha$-helix),
in which all the amino acids occupy bounded states.

\item[II.] partially folded states of the polypeptide (phase
co-existence), in which the core of $\lambda$ amino acids of the
polypeptide occupy bounded states, and $n-\lambda$ boundary amino
acids are in unbounded states.

\item[III.] completely unfolded state of a polypeptide (random coil), in which all the amino acids
are in unbounded states.

\item[IV.] phase mixing, in which two or more
fragments of a polypeptide are in an $\alpha$-helix state, while the
amino acids between the fragments are in the random coil state.
\end{itemize}

With the assumptions outlined above and assuming the polypeptide to
consist of $n$ identical amino acids the partition function
(\ref{Z_step2}) of the system can be rewritten as follows:

\begin{eqnarray}
\nonumber {\mathbb Z}&=&A\cdot
B(kT)\cdot(kT)^{3N-3-\frac{l_s}{2}}\left[\beta
Z_{b}^{n-1}Z_{u}+\beta\sum_{i=1}^{n-4}(i+1)Z_{b}^{n-i-1}Z_{u}^{i+1}+Z_{u}^n+\right.\\
&&\left.+\sum_{i=2}^{(n-3)/2}\beta^i\sum_{k=i}^{n-i-3}\frac{(k-1)!(n-k-3)!}{i!(i-1)!(k-i)!(n-k-i-3)!}Z_b^{k+3i}Z_u^{n-k-3i}\right]
\label{Zhc}
\end{eqnarray}

\noindent Here the first and the third terms in the square brackets
describe the partition function of the polypeptide in the
$\alpha$-helix and in the random coil phases respectively, while the
second term in the square brackets accounts for situation of the
phase co-existence. The summation in the second term in (\ref{Zhc})
is performed up to $n-4$, because the shortest $\alpha$-helix
consists of 4 amino acids. The last term in the square brackets
accounts for the polypeptide conformations in which a number of
amino acids being in the helix conformation are separated by amino
acids being in the random coil conformation. The first summation in
this term goes over the separated helical fragments of the
polypeptide, while the second summation goes over individual amino
acids in the corresponding fragment. Polypeptide conformations with
two or more helical fragments are energetically unfavorable. This
fact is discussed in our following paper
\cite{Yakubovich07_following}. As shown in the following paper
\cite{Yakubovich07_following} the contribution to the partition
function represented by the fourth term in the square brackets in
Eq.~(\ref{Zhc}) is significantly small when compared to the first
three terms, for polypeptides containing less than 100 of amino
acids. Therefore, it can be omitted in the construction of the
partition function. $Z_b$ and $Z_u$ are the contributions to the
partition function from a single amino acid being in the bounded or
unbounded states respectively, they read as:

\begin{eqnarray}
\label{Zb} Z_b&=&\int_{-\pi}^{\pi }\int_{-\pi}^{ \pi}
\exp\left({-\frac{\epsilon^{(b)}(\varphi,\psi)}{kT}}\right){\rm
d}\varphi{\rm d}\psi\\
\label{Zu} Z_u&=&\int_{-\pi}^{\pi }\int_{-\pi}^{ \pi}
\exp\left({-\frac{\epsilon^{(u)}(\varphi,\psi)}{kT}}\right){\rm
d}\varphi{\rm d}\psi\\
\label{Beta} \beta&=&\left(\int_{-\pi}^{\pi }\int_{-\pi}^{ \pi}
\exp\left({-\frac{\epsilon^{(b)}(\varphi,\psi)+\epsilon^{(u)}(\varphi,\psi)}{kT}}\right){\rm
d}\varphi{\rm d}\psi\right)^3,
\end{eqnarray}

\noindent where $\epsilon^{(b)}(\varphi,\psi)$ and
$\epsilon^{(u)}(\varphi,\psi)$ are the potential energies of a
single amino acid being in the bounded or in the unbounded states
respectively calculated versus the twisting degrees of freedom
$\varphi$ and $\psi$.  $\beta$ is a factor accounting for the
entropy loss of the helix initiation. Substituting (\ref{Zb}),
(\ref{Zu}) and (\ref{Beta}) into equation (\ref{Zhc}) one obtains
the final expression for the partition function of polypeptide
undergoing an $\alpha$-helix$\leftrightarrow$random coil phase
transition. This result can be used for the evaluation of all
thermodynamical characteristics of the system.

$\epsilon^{(b)}(\varphi,\psi)$ and $\epsilon^{(u)}(\varphi,\psi)$
determine the partition function of polypeptide. These quantities
can be calculated on the basis of {\it ab initio} DFT, combined with
molecular mechanics theories as demonstrated in
\cite{Yakubovich06a,Yakubovich06a_EPN} and in the following paper
\cite{Yakubovich07_following}.

\section{Thermodynamical characteristics of a polypeptide chain}

\begin{figure}[h]
\includegraphics[scale=0.7,clip]{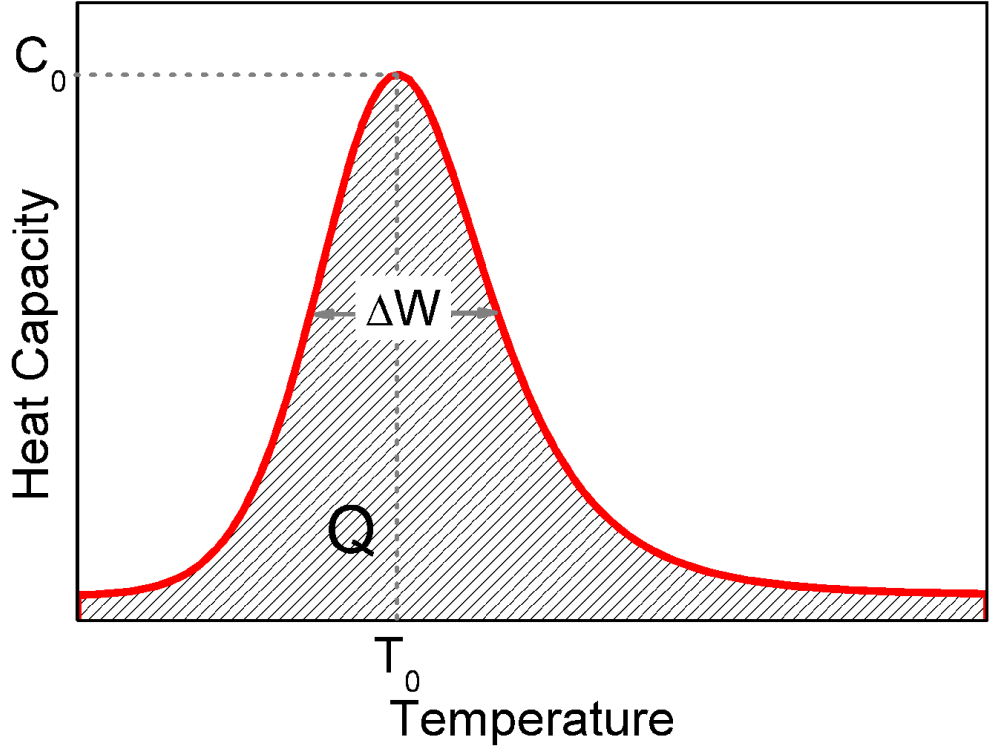}
\caption{Temperature dependence of the heat capacity for a system
experiencing a phase transition.} \label{fg:HeatCapacity_parametric}
\end{figure}

The first order phase transition is characterized by an abrupt
change of the internal energy of the system with respect to its
temperature. In the first order phase transition the system either
absorbs or releases a fixed amount of energy while heat capacity as
a function of temperature has a sharp peak
\cite{Ptizin_book,Landau05} (see Fig.
\ref{fg:HeatCapacity_parametric}).

The peak in the heat capacity is characterized by the transition
temperature $T_{0}$, the maximal value of the heat capacity $C_0$,
the temperature range of the phase transition $\Delta W$ and the
specific heat $Q$, which is also referred as the latent heat of the
phase transition (see Fig. \ref{fg:HeatCapacity_parametric}).

All these quantities can be calculated if the dependence of the heat
capacity on temperature is known. The temperature dependence of the
heat capacity is defined by the partition function as follows
\cite{Landau05}:

\begin{equation}
C(T)=kT\frac{\partial^2T\ln {\mathbb Z}}{\partial T^2}. \label{C(T)}
\end{equation}

\noindent The characteristics of the phase transition are determined
by the following equations:

\begin{eqnarray}
\label{T0_gen}
\left.\frac{dC(T)}{dT}\right|_{T=T_0}=0\\
\label{C0_gen} C_{0}=C(T_0)\\
\label{dW_gen} C(T_0\pm\Delta W)=\frac{C_0}{2}\\
\label{Q_gen} Q=\int_{0}^{\infty}C(T){\rm d}T.
\end{eqnarray}

\noindent Unfortunately it is not possible to obtain analytical
expressions for $T_{0}$, $C_0$, $\Delta W$ and $Q$ with partition
function defined in (\ref{Zhc}) because the integrals in (\ref{Zb})
and (\ref{Zu}) can not be treated analytically. However, the
qualitative behavior of these quantities can be understood if one
assumes that all conformational states of a polypeptide in a certain
phase have the same energy. This model is usually referred to in
literature as the two-energy-level model
\cite{Prabhu05,Yakubovich06a,Yakubovich06a_EPN} and it turns out to
be very useful for the qualitative analysis of the phase transitions
in polypeptide chains. If one considers the phase transition between
two such phases, the partition function can then be constructed as
follows:

\begin{equation}
{\mathbb Z} \approx {\mathbb Z_0}\left[1+A\frac{\eta_2}{\eta_1}e^{-
\frac{\Delta E}{kT}}\right], \label{Z2sys}
\end{equation}

\noindent where ${\mathbb Z_0}$ is the partition function of the
system in the first phase, $\Delta E=E_2-E_1$ is the energy
difference between the states of the polypeptide in two different
phases, $\eta_1$ and $\eta_2$ are the numbers of isomeric states of
the polypeptide in the first and in the second phases respectively.
They can also be considered as the population of the two phases.
$A=A_2/A_1$ is the coefficient depending on masses, specific
volumes, normal vibration modes frequencies and momenta of inertia
of the polypeptide in the two phases. Substituting equation
(\ref{Z2sys}) into equation (\ref{C(T)}) one obtains the expression
for the heat capacity in the framework of the two-energy-level
model:

\begin{equation}
C(T)= \frac{A\frac{\eta_2}{\eta_1}\Delta E^2e^{-\left(\frac{\Delta
E}{kT}\right)}}{kT^2\left(1+A\frac{\eta_2}{\eta_1}e^{-\left(\frac{\Delta
E}{kT}\right)}\right)^2}. \label{C(T)2lev}
\end{equation}

\noindent Substituting equation (\ref{C(T)2lev}) into equations
(\ref{T0_gen})-(\ref{Q_gen}) and solving them one obtains the
expressions for $T_{0}$, $C_{0}$, $\Delta W$ and $Q$, which read as:

\begin{eqnarray}
\label{parametricT0}
T_{0}&\approx&\frac{{\Delta E}}{k\ln\left(A\frac{\eta_2}{\eta_1}\right)}=\frac{{\Delta E}}{\Delta S},\\
\label{parametricC0}
C_{0}&\approx&\frac{k}{4}\left[\ln\left(A\frac{\eta_2}{\eta_1}\right)\right]^2=\frac{\Delta S^2}{4k},\\
\label{parametricDW}\Delta
W&\approx&\sqrt{\frac{64\ln2}{\pi}}\frac{\Delta
E}{k\left[\ln\left(A\frac{\eta_2}{\eta_1}\right)\right]^2}=\sqrt{\frac{64\ln2}{\pi}}\frac{
k\Delta E}{\Delta S^2},\\
\label{parametricQ} Q&=&\int C(T){\rm d}T=\Delta E.
\end{eqnarray}

\noindent Here $\Delta S=k\ln{A\eta_2}-k\ln{\eta_1}$ is the entropy
change in the system and $M$ is the mass of a single polypeptide.
$\Delta S$ and $\Delta E$ are the major thermodynamical parameters
in the considered problem, since they determine the behavior of the
phase transition characteristics. From equations
(\ref{parametricT0})-(\ref{parametricDW}) follows, that
$T_0\sim\frac{{\Delta E}}{\Delta S}$, $C_0\sim\Delta S^2$,
$Q\sim\Delta E$ and $\Delta W\sim\frac{\Delta E}{\Delta S^2}$.

The numerical calculation and analysis of various thermodynamical
characteristics such as the latent heat or the heat capacity is done
in the following paper \cite{Yakubovich07_following}.

\section{Conclusion}
\label{conclusion}

In the present paper a novel {\it ab initio} theoretical method for
treating the $\alpha$-helix$\leftrightarrow$random coil phase
transition in polypeptide chains is introduced. The suggested method
is based on the construction of a parameter-free partition function
for a system undergoing a first order phase transition. All the
necessary information for the construction of such a partition
function can be calculated on the basis of {\it ab initio} DFT,
combined with molecular mechanics theories (see results of numerical
simulations in the following paper \cite{Yakubovich07_following}).

The suggested method is considered as an efficient alternative to
the existing theoretical approaches for the study of helix-coil
transition in polypeptides since it does not contain any model
parameters. It gives a universal recipe for statistical mechanics
description of complex molecular systems. The partition function of
polypeptide is written with a minimum number of assumptions about
the system which makes our method much more general and universal in
comparison with other theoretical approaches.

In the present paper we introduced novel theoretical method for the
study of $\alpha$-helix$\leftrightarrow$random coil phase transition
in polypeptides. In the following paper
\cite{Yakubovich07_following} we report the results of numerical
simulations of this process obtained within the framework of the
suggested model.

\section{Acknowledgments}
We acknowledge support of this work by the NoE EXCELL, by INTAS
under the grant 03-51-6170. We are grateful to Ms. Stephanie Lo for
critical reading of the manuscript and several suggestions for
improvement.

\bibliography{journals_short,PhaseTrans_in_Polypeptides}

\end{document}